\documentclass[epj, final]{svjour}

\usepackage[pdftex]{graphicx}
\usepackage{subfig}

\usepackage[sort&compress]{natbib}
\bibpunct{[}{]}{,}{n}{}{}

\usepackage{amsmath}
\usepackage{amssymb}
\usepackage{bbm}
\usepackage{braket}
\usepackage{mathtools}

\newcommand{\fig}[1]{Fig.~\ref{fig:#1}}
\newcommand{\eq}[1]{(\ref{eq:#1})}
\newcommand{\lr}[1]{\ensuremath{\left( #1 \right)}}
\newcommand{\dg}{+}

\renewcommand{\d}{\mathrm{d}}

\newcommand{\I}{\mathrm{i}}

\newcommand{\sg}{\sigma}
\newcommand{\Sg}{\Sigma}

\newcommand{\Dl}{\Delta}

\newcommand{\veps}{\varepsilon}
\newcommand{\lb}{\lambda}

\newcommand{\ovs}{\overset}

\newcommand{\Abs}[1]{\ensuremath{\left| #1 \right|}}

\renewcommand{\Im}[1]{\ensuremath{\mathrm{Im} (#1)}}

\newcommand{\Tr}[1]{\ensuremath{\mathrm{Tr}(#1)}}

\newcommand{\dE}{\ensuremath{\d E\,}}

\begin{document}
\title{Statistical model for the effects of phase and momentum randomization on electron transport}
\titlerunning{Statistical model for the effects of phase and momentum randomization on electron transport}
\author{Thomas Stegmann\inst{1}\fnmsep\thanks{\email{thomas.stegmann@uni-due.de}} \and Mat\'{i}as Zilly\inst{1} \and Orsolya Ujs\'{a}ghy\inst{2} \and
  Dietrich E. Wolf\inst{1}} 
\institute{Department of Physics and CeNIDE, University of Duisburg-Essen, 47048 Duisburg, Germany \and Department of Theoretical Physics, Budapest
  University of Technology and Economics, 1521 Budapest, Hungary}
\authorrunning{T. Stegmann et al.}
\date{Received 27 April 2012\\Published 30 July 2012} 

\abstract{A simple statistical model for the effects of dephasing on electron transport in one-dimensional quantum systems is introduced, which allows
  to adjust the degree of phase and momentum randomization independently. Hence, the model is able to describe the transport in an intermediate regime
  between classical and quantum transport. The model is based on B\"uttiker's approach using fictitious reservoirs for the dephasing effects. However,
  in contrast to other models, at the fictitious reservoirs complete phase randomization is assumed, which effectively divides the system into smaller
  coherent subsystems, and an ensemble average over randomly distributed dephasing reservoirs is calculated. This approach reduces not only the
  computation time but allows also to gain new insight into system properties. In this way, after deriving an efficient formula for the
  disorder-averaged resistance of a tight-binding chain, it is shown that the dephasing-driven transition from localized-exponential to ohmic-linear
  behavior is not affected by the degree of momentum randomizing dephasing.}
\maketitle


\section{Introduction}
Electron transport through nanosized systems takes place in an intermediate regime between classic and quantum transport because the characteristic
lengths, system size $L$, phase coherence length $l_\phi$, and mean free path $l_m$ have the same order of magnitude. By using these lengths, the
transport can be classified roughly into the following re\-gimes. A system with phase and momentum randomization $l_\phi \sim l_m < L$, for which the
resistance is ohmic, i.e. it increases linearly with the system size. Coherent transport through a disordered system $l_m < l_\phi \sim L$, for which
the resistance increases exponentially with the system size. A homogeneous system without momentum randomization $l_\phi \lesssim l_m \sim L$, for
which the conductor is ballistic, i.e. its resistance is length independent. The interplay of these completely different transport regimes makes
nanosized systems very promising for novel electronic devices. Hence the need for a theoretical description, which covers all these regimes, is
obvious.

Basically, the nonequilibrium Green's function (NEGF) approach for the transport through quantum systems \cite{Caroli1971, Datta1997, Datta2005}
allows to include arbitrary interactions by self-energies, which cause phase and momentum randomization. By using the first order self-consistent Born
approximation for the self-energies, Datta proposed a model \cite{Golizadeh-Mojarad2007}, which allows to adjust the degree of phase and momentum
randomization independently. However, its application to nanosized systems involves enormous computational efforts. Hence, simple models are required
to include the effects of dephasing in nanosized systems.

B\"uttiker proposed to use fictitious reservoirs as a dephasing model, where the phase of the electrons is lost by absorption and reinjection
\cite{Buettiker1986_1, Buettiker1991}. These fictitious reservoirs are used nowadays in several models \cite{Amato1990, Maschke1994, Knittel1999,
  Li2002, Yu2002, Roy2007, Nozaki2012}. Note, that this apparently phenomenological approach of fictitious reservoirs can be justified from a
microscopic theory under appropriate approximations \cite{McLennan1991, Hershfield1991}.

Other models include the dephasing effects by stochastic absorption through an attenuating factor \cite{Joshi2000} or by random phase factors
\cite{Pala2004, Zheng2006}. However, these models are still controversially discussed \cite{Bandopadhyay2010} and most of them are defined only in the limit of
completely momentum randomizing or completely momentum conserving dephasing.

In this paper we present a simple statistical model for the effects of dephasing on electron transport in one-dimensional quantum systems which allows
to adjust the degree of phase and momentum randomization independently. The model is based on B\"uttiker's approach but in contrast to other models,
complete loss of the phase at each dephasing reservoir is assumed, and afterwards an ensemble average of the quantity of interest (e.g. resistance or
conductance) over several spatial configurations of dephasing reservoirs is calculated. The model is discussed in detail in Section~\ref{sec:Model}
and applied to linear tight-binding chains in Section~\ref{sec:Application}. Concluding remarks can be found in Section~\ref{sec:conclusion}.

The model is an extension of our previous work \cite{Zilly2009, ZillyPhD2010}, where the limit of completely momentum randomizing dephasing has been
discussed. In this limit the model has been used successfully to explain several conductance measurements of DNA chains \cite{Zilly2010}. Moreover, by
applying it to the 1D Anderson model with arbitrary uncorrelated disorder, the decoherence induced conductivity has been studied and even-order
generalized Lyapunov exponents have been calculated, which were used to approximate the localization length \cite{Zilly2012}.

\begin{figure}[b]
  \subfloat{\hspace*{5ex}\includegraphics{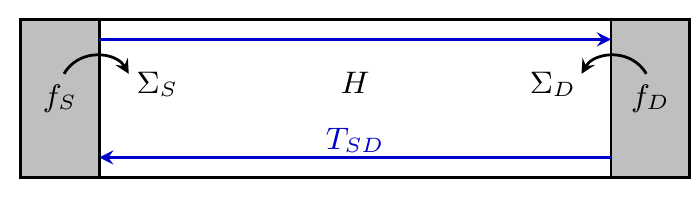}(a)}\\
  \subfloat{\hspace*{5ex}\includegraphics{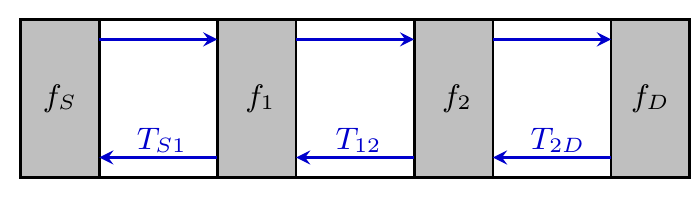}(b)}\\
  \subfloat{\hspace*{5ex}\includegraphics{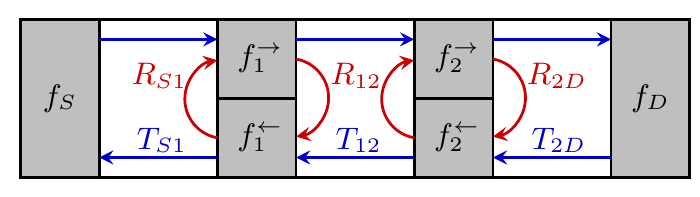}(c)}
  \caption{(Color online) (a) Coherent transport through a one-dimensional quantum system. Fermi distributions $f_{S/D}$ are assumed for the source
    and the drain reservoir. (b) Dephasing is introduced by fictitious dephasing reservoirs, where the phase and the momentum of the electrons are
    completely randomized. The dephasing reservoirs divide the system into smaller coherent subsystems and are characterized by an energy distribution
    function $f_k$. (c) Momentum conserving dephasing is obtained if two energy distribution functions per dephasing reservoir are defined, which
    allow to distinguish between the electrons with positive $f_k^\rightarrow$ and negative momentum $f_k^\leftarrow$.}
  \label{fig:1}
\end{figure}

\section{Model} \label{sec:Model}
We consider a one-dimensional quantum system with the single-particle Hamiltonian $H$. The system is coupled to two reservoirs, which act as the
source and the drain of electrons, see \fig{1}(a). The coherent transport from the source to the drain is characterized by a transmission function,
which is calculated within the NEGF approach \cite{Datta2005}
\begin{equation} \label{eq:transmission}
  T_{SD}= 4 \Tr{\Im{\Sg_S}G\Im{\Sg_D}G^{\dg}},
\end{equation}
where
\begin{equation}
  G= \lr{E-H-\Sg_S-\Sg_D}^{-1}
\end{equation}
is the Green function of the system. The self-energies $\Sg_{S/D}$ represent the influence of the source respectively the drain reservoir on the
system. The current is calculated by the Landauer formula \cite{Datta2005}
\begin{equation}
   I= \frac{e}{h} \int \dE T_{SD} \lr{f_S-f_D}
\end{equation}
where $f_{S/D}=f(E-\mu_{S/D})$ are the Fermi distributions of the reservoirs with the electrochemical potential $\mu_{S/D}$.

\subsection{Momentum randomizing dephasing}
At first, we briefly summarize our statistical model for the effects of momentum randomizing dephasing \cite{Zilly2009, ZillyPhD2010}. 

Dephasing is introduced by a spatial configuration of fictitious \textit{dephasing reservoirs}, where the phase of the electrons as well as their
momentum are completely randomized, see \fig{1}(b). Thus, these dephasing reservoirs divide the one-dimensional system into smaller coherent
subsystems with the transmission $T_{k,k+1}$. An energy distribution function $f_k$ is assigned to each of the $k=1,2 \hdots N_d$ dephasing
reservoirs, which can be calculated by using the conservation of the energy resolved current at the fictitious reservoirs
\begin{equation} \label{eq:ratemr}
  T_{k-1,k}\lr{f_{k}-f_{k-1}} -T_{k,k+1}\lr{f_{k+1}-f_{k}}\ovs{!}{=}0.
\end{equation}
The solution of this system of $N_d$ linear equations for the energy distribution function of the last dephasing reservoir is \cite{Zilly2009,
  ZillyPhD2010}
\begin{equation}
  f_{N_d}= \frac{\frac{1}{T_{N_d,D}}f_S +\sum_{k=0}^{N_d-1}\frac{1}{T_{k, k+1}}f_D}{\sum_{k=0}^{N_d}\frac{1}{T_{k,k+1}}}.
\end{equation}
This allows the calculation of the current through the system
\begin{equation}
  I= \frac{e}{h}\int \dE T_{N_d,D}\lr{f_{N_d}-f_D}= \frac{e}{h} \int \dE \frac{f_S-f_D}{\sum_{k=0}^{N_d}\frac{1}{T_{k,k+1}}}.
\end{equation}
Hence, at an infinitesimal bias voltage $\mu_S \to \mu_D$, the total resistance measured in multiples of $h/e^2$
\begin{equation} \label{eq:rmr}
  R= \sum_{k=0}^{N_d}\frac{1}{T_{k,k+1}}
\end{equation}
is given by the sum of the subsystem resistances, which are determined by the coherent quantum transport between neighboring reservoirs. Moreover, we
do not only consider a single fixed spatial configuration of dephasing reservoirs but calculate the ensemble average of the total resistance $\{R\}$
over the dephasing configurations. Note also that by assuming an infinitesimal bias voltage the above used conservation of the energy resolved current
is equivalent to the conservation of the total current.

\subsection{Momentum conserving dephasing}
Momentum conserving dephasing is obtained in our model, if two energy distribution functions per dephasing reservoir are defined, see \fig{1}(c). The
$f_k^\rightarrow$ and $f_k^\leftarrow$ allow to attribute to every electron a definite sign of the momentum. The absence of energy relaxation then
ensures momentum conservation during the dephasing process. The current conservation constraint causes $2N_d$ linear equations
\begin{align} \label{eq:ratemc}
  T_{k-1,k}f_{k-1}^\rightarrow -T_{k,k+1}f_{k}^\rightarrow -R_{k,k+1}f_{k}^\rightarrow +R_{k-1,k}f_{k}^\leftarrow &\ovs{!}{=}0, \notag\\
  T_{k,k+1}f_{k+1}^\leftarrow -T_{k-1,k}f_{k}^\leftarrow -R_{k-1,k}f_{k}^\leftarrow +R_{k,k+1}f_{k}^\rightarrow &\ovs{!}{=}0,
\end{align}
where $R_{i,j}=1-T_{i,j}$. The energy distribution function of the last dephasing reservoir reads
\begin{equation}
  f_{N_d}^\rightarrow= \frac{\frac{1}{T_{N_d,D}}f_S +\sum_{k=0}^{N_d-1}\frac{1}{T_{k,k+1}}f_D -N_d f_D}{\sum_{k=0}^{N_d}\frac{1}{T_{k,k+1}} -N_d}
\end{equation}
and the total resistance at an infinitesimal bias voltage
\begin{equation} \label{eq:rmc}
  R= \sum_{k=0}^{N_d}\frac{1}{T_{k,k+1}}-N_d.
\end{equation}
Hence, for momentum conserving dephasing, the sum of the subsystem resistances is reduced by a constant contact resistance $R_c=1$ for each of the fictitious reservoirs.

If momentum randomizing as well as momentum conserving dephasing exist in the same system, \eq{ratemc} can still be used for the momentum conserving
reservoirs, if one sets $f_{k}^\rightarrow=f_{k}^\leftarrow=f_k$ for the momentum randomizing dephasing reservoirs and uses for these reservoirs a
modified version of \eq{ratemr} $ T_{k-1,k}\lr{f_{k}-f_{k-1}^\rightarrow} -T_{k,k+1}\lr{f_{k+1}^\leftarrow-f_{k}}=0$.

\section{Application to linear tight-binding chains} \label{sec:Application}
In the following we consider linear tight-binding chains of $N$ sites with a single energy level $\veps_k$ per site and coupling $t$ between nearest neighbors
\begin{equation}
  H= \sum_{k=1}^N \veps_k c_k^\dg c_k +t\sum_{k=1}^{N-1} \lr{c_k^\dg c_{k+1} +c_{k+1}^\dg c_k}.
\end{equation}
To simplify the notation we take $t=1$ as our energy unit and the lattice spacing $a=1$ as our length unit. The reservoirs are modeled as
semi-infinite chains which cause the self-energy (see for example \cite{ZillyPhD2010})
\begin{equation} \label{eq:selfenergy}
  \Sg=
  \begin{cases}
    \frac{E}{2} +\frac{1}{2} \sqrt{E^2-4} &\quad \text{for }E \le -2,\\
    \frac{E}{2} -\frac{\I}{2} \sqrt{4-E^2} &\quad \text{for }\Abs{E} < 2,\\
    \frac{E}{2} -\frac{1}{2} \sqrt{E^2-4} &\quad \text{for }E \ge +2
  \end{cases}
\end{equation}
on the first and last site of the chain. In the following, only the energy band $\Abs{E}<2$ is considered because for $\Abs{E}>2$ the imaginary part
of the self-energy vanishes and hence $T=0$. The coherent transmission through a chain of length $N$ is calculated recursively 
by using the same method as in \cite{Zilly2012}
\begin{equation}\label{eq:transrec}
  \frac{1}{T_N(E)}= \frac{\Abs{r_N +\Sg^2 s_{N-1} -\Sg (r_{N-1} +s_N)}^2}{4 \Im{\Sg}^2} 
\end{equation}
with the polynomials
\begin{align} \label{eq:polyrec}
  &r_j=(E-\veps_j)r_{j-1}-r_{j-2}, \quad &&s_j=(E-\veps_j)s_{j-1}-s_{j-2}, \notag\\
  &r_0=1, \quad &&s_1=1,\notag\\
  &r_{-1}=0, \quad &&s_0=0.
\end{align}

\begin{figure}[t]
  \centering
  \includegraphics[scale=0.39]{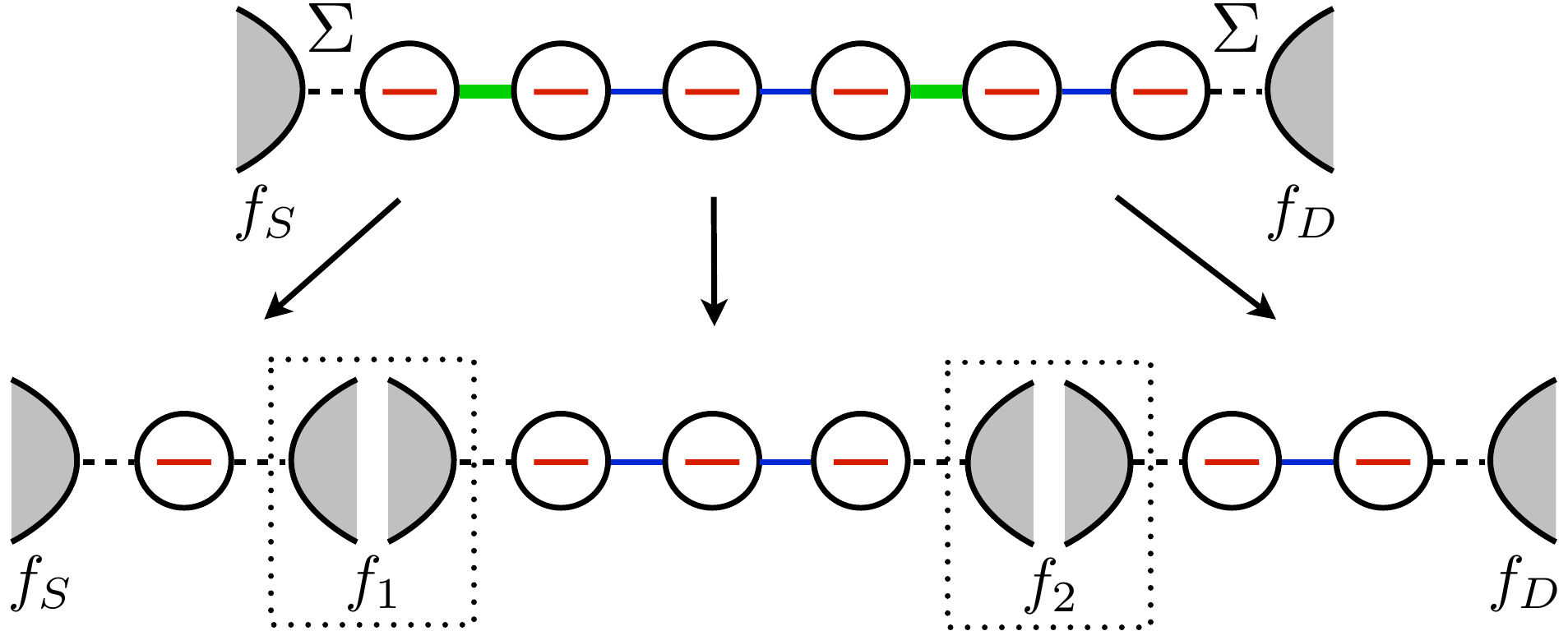}
  \caption{(Color online) A dephasing configuration is generated by replacing randomly chosen bonds of the tight-binding chain with fictitious
    dephasing reservoirs.}
  \label{fig:2}
\end{figure}

\begin{figure}[t]
  \centering
  \includegraphics{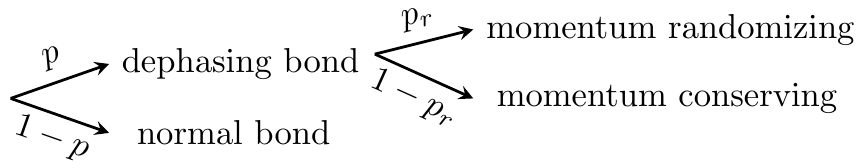}
  \caption{Decision tree for generating dephasing configurations.}
  \label{fig:3}
\end{figure}

A dephasing configuration is generated by replacing each bond of the chain by a dephasing reservoir with probability $p$, see \fig{2}. These
dephasing reservoirs are momentum randomizing with probability $p_r$ and hence momentum conserving with probability $1-p_r$, see \fig{3}.

The phase coherence length $l_\phi$ is defined by the average length of the coherent subsystems
\begin{equation} \label{eq:lphi}
  l_\phi= \frac{N}{1+(N-1)p}.
\end{equation}
The average distance between momentum randomizing dephasing reservoirs reads
\begin{equation}  \label{eq:lmdph}
  l_m^{dph}= \frac{N}{1+(N-1)pp_r}.
\end{equation}

If only $N_d^c$ of the $N_d$ dephasing reservoirs are assumed as momentum conserving, the ensemble average of the resistance over all dephasing
configurations is given by
\begin{equation} \label{eq:Rav}
  \bigl\{ R \bigr\}= \biggl\{ \sum_{k=0}^{N_d} \frac{1}{T_{k,k+1}} \biggr\} -(N-1)p(1-p_r)
\end{equation}
because $\{N_d^c\}= (N-1)p(1-p_r)$.

The additional resistance due to momentum randomizing dephasing
\begin{equation} \label{eq:dR}
  \left \{ \Dl R \right\}\equiv \{ R(p_r)\}-\{ R(p_r=0)\}= (N-1)pp_r
\end{equation}
increases linearly with the chain length and both dephasing probabilities. 

Regarding a fixed dephasing configuration as a series of incoherently coupled tunneling barrieres, its resistance in the limit of completely momentum
randomizing dephasing \eq{rmr} and completely momentum conserving dephasing \eq{rmc} corresponds to the results in \cite{Buettiker1991, Yu2002,
  Bandopadhyay2010}. However, by ensemble averaging our statistical dephasing model provides a simple formula \eq{dR} for the additional resistance
due to momentum randomizing dephasing, which is also valid for partial decoherence $p<1$ and partial momentum randomization $0<p_r<1$.

\subsection{Homogeneous chains}
Homogeneous chains $\veps_k=0$ are considered, which show perfect transmission $T=1$ if $\Abs{E}<2$. The ensemble averaged resistance
\begin{align} \label{eq:Ravh}
  \{ R \}&= \{N_d+1\} -(N-1)p(1-p_r)\notag \\
  &= 1 +(N-1)pp_r= N/l_m
\end{align}
is given by the ratio of the chain length and the mean free path since in homogeneous chains momentum randomization is caused only by the dephasing
reservoirs and hence $l_m=l_m^{dph}$. Obviously, momentum conserving dephasing $p_r=0$ does not cause an additional resistance and retains a ballistic
conductor $\{R\}=1$, whereas momentum randomizing dephasing $p_r>0$ leads to an ohmic resistance $\{R\} \propto N$.

\subsection{Disordered chains}
In this section the onsite energies are considered as distributed independently according to a probability density $w(\veps)$ with mean value 0 and
variance $\sg^2$.

The onsite disorder causes momentum randomization, which for large $N$ can be estimated by the exponential decrease of the coherent transmission $T
\propto \exp(-N/\lb)$. The mean free path $l_m$ is then given by summing up the spatial rates of the two momentum randomization processes
\begin{equation} \label{eq:lm} 
  \frac{1}{l_m}= \frac{1}{l_m^{dph}} +\frac{1}{\lb}.
\end{equation}

\begin{figure}[t]
  \centering
  \includegraphics{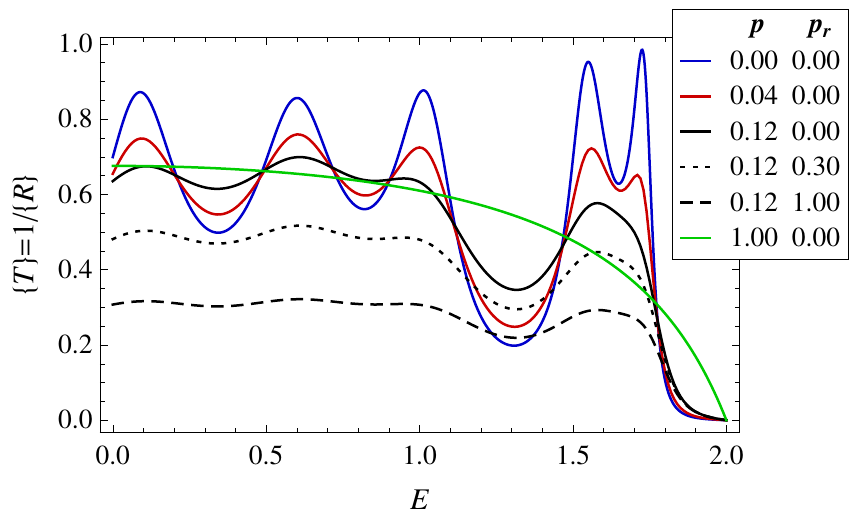}
  \caption{(Color online) The coherent transmission through a chain of 15 sites with a given disorder configuration shows oscillations due to the
    interference between the disordered sites. When dephasing $p>0$ is introduced, the interference is reduced and hence the oscillations vanish. If
    the dephasing is momentum conserving $p_r=0$ only the oscillations are averaged out, whereas the transmission itself is reduced if the dephasing
    is momentum randomizing $p_r>0$. This clearly indicates the additional resistance.}
  \label{fig:4}
\end{figure}

The energy resolved transmission of a fixed disorder configuration in \fig{4} shows strong oscillations due to interference, if the transport is
coherent. When dephasing is introduced, $p>0$, these oscillations vanish because the interference between the disordered sites is no longer possible. If
the dephasing is momentum conserving $p_r=0$, only the oscillations are averaged out, whereas the transmission itself is reduced, when the dephasing is
momentum randomizing $p_r>0$. This property of the transmission function, which is also reported in \cite{Yu2002, Zheng2006, Golizadeh-Mojarad2007},
clearly indicates the additional resistance due to momentum randomizing dephasing.

Now, disorder averages are studied. The disorder-aver\-aged resistance of a coherent chain of $j$ sites is calculated recursively in the same ways as in
\cite{Zilly2012}
\begin{align} \label{eq:rdrec}
  \left\langle \frac{1}{T_j} \right \rangle &\equiv \int \d \, \veps_1 \hdots \d \veps_j \frac{1}{T_j} \prod_{i=1}^j w(\veps_i) \notag\\
  &= \frac{P_j+2P_{j-1}\lr{1-E^2}+P_{j-2}+E^2 R_j +2}{4-E^2}
\end{align}
with the recursion relations
\begin{subequations}
  \begin{align}
    P_j &\equiv \int \d \veps_1 \hdots \d\veps_j \, r_j^2 \prod_{i=1}^j w(\veps_i) \notag\\
    &=\lr{E^2 + \sg^2}P_{j-1} -2E Q_{j-1}+P_{j-2},\\
    Q_j &\equiv \int \d \veps_1 \hdots \d\veps_j \, r_jr_{j-1} \prod_{i=1}^j w(\veps_i) \notag\\
    &=E P_{j-1} -Q_{j-1},\\
    R_j&\equiv \int \d \veps_1 \hdots \d\veps_j \, r_j s_{j-1} \prod_{i=1}^j w(\veps_i) \notag\\ 
    &=EQ_{j-1}-R_{j-1}-1
  \end{align}
\end{subequations}
and the initial conditions $P_0=1, \: P_{-1}= Q_0= R_1=0$. Note that the disorder-averaged resistance depends only on the variance of the probability
density $w$, higher moments do not enter.

\begin{figure}[t]
  \centering
  \includegraphics{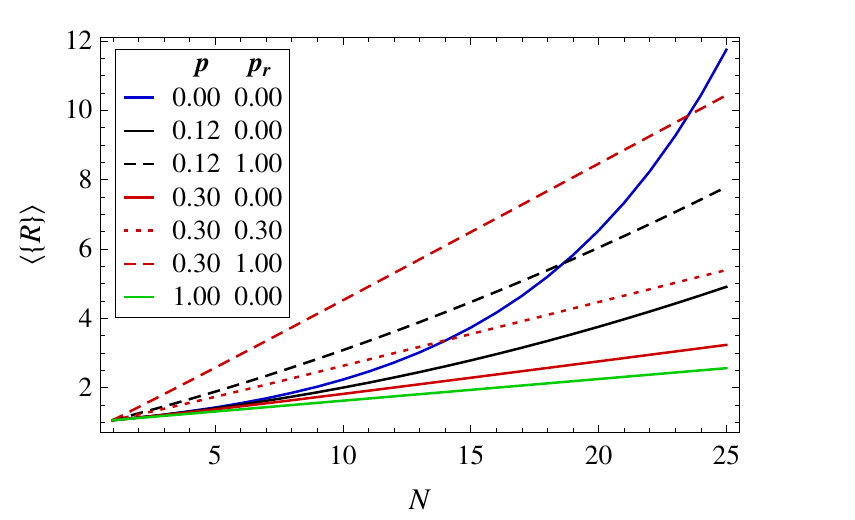}
  \caption{(Color online) Length dependent disorder-averaged resistance in the band-center $E=0$ ($\sg=0.5$). The resistance transits from
    localized-exponential to ohmic-linear behavior, if the dephasing probability is sufficiently increased $p>p^*>0$. Momentum conserving dephasing
    $p_r=0$ always reduces the resistance as the fictitious reservoirs do not cause extra resistance but reduce the tunneling distance. Thus, the
    resistance is minimal if every bond is replaced by a momentum conserving dephasing reservoir. If the dephasing is momentum randomizing, the
    resistance depends crucially on the chain length, the disorder strength, the dephasing probabilities, and the energy.}
  \label{fig:5}
\end{figure}

If dephasing bonds are introduced with probability $p$ in a chain of $N$ sites, the coherent chain appears obviously with probability $(1-p)^{N-1}$. A
subsystem with length $j<N$ appears with probability $p(1-p)^{j-1}$ at one of the two chain ends because it is separated by only one dephasing bond
from the remaining sites. However, inside the chain it is separated by two dephasing bonds and appears with probability $p^2(1-p)^{j-1}$ at one of the
$N-1-j$ possible positions. Hence, the average number of subsystems with length $j$ in a chain of $N$ sites is given by
\begin{equation} \label{eq:lav}
  u_j=
  \begin{cases}
    \lr{1-p}^{j-1} \bigr( 2p +\lr{N-1-j}p^2 \bigl) \quad &\text{for } j<N,\\
    \lr{1-p}^{N-1} \quad &\text{for } j=N.
  \end{cases}
\end{equation}
The averaged resistance reads
\begin{equation} \label{eq:Ravd}
  \left\langle \{R \} \right\rangle= \sum_{j=1}^N u_j \left\langle\frac{1}{T_j} \right\rangle -\lr{N-1}p(1-p_r).
\end{equation}

\begin{figure}[t]
  \centering
  \includegraphics{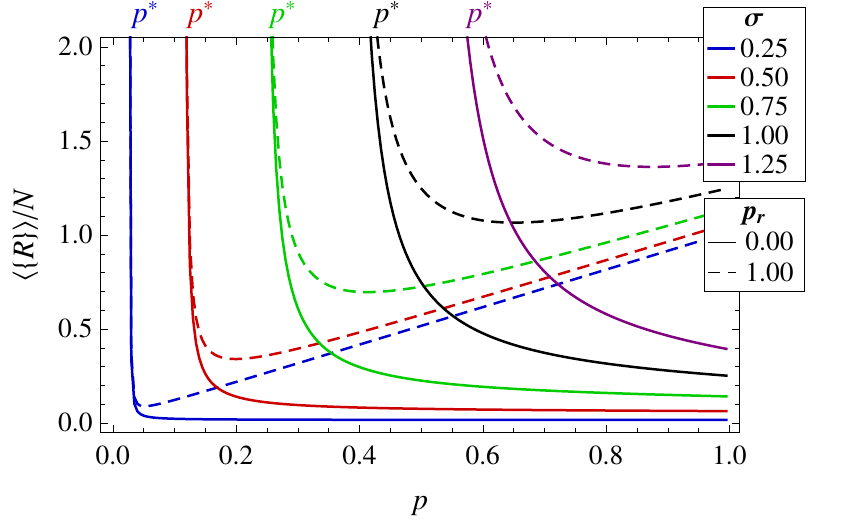}
  \caption{(Color online) Average resistance per length of a chain with $N=1000$ sites in the band-center $E=0$. The critical dephasing probability
    $p^*$ indicates the transition from localized-exponential to ohmic-linear behavior.}
  \label{fig:6}
\end{figure}

\begin{figure}[t]
  \centering
  \includegraphics{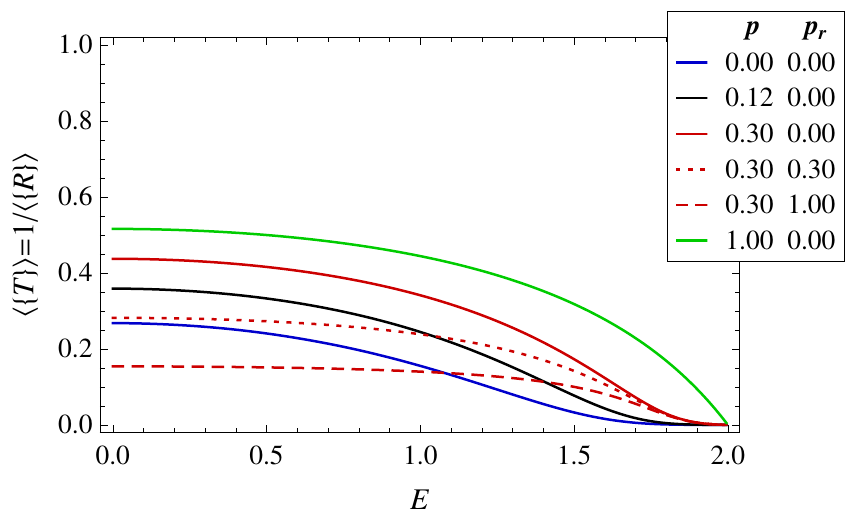}
  \caption{(Color online) Disorder-averaged transmission of the 15-site chain ($\sg=0.5$). Oscillations as in \fig{4} do not appear due to the ensemble average
    over disorder configurations. The transmission is always increased by momentum conserving dephasing $p_r=0$, whereas it depends crucially on the
    chain length, the disorder strength, the dephasing probabilities, and the energy, if the dephasing is momentum randomizing $p_r>0$, cf. \fig{5}.}
  \label{fig:7}
\end{figure}

\fig{5} shows the resistance in the band-center $E=0$ as a function of the chain length $N$. Because of the disorder-induced ($\sg=0.5$) quantum
localization, the coherent resistance $p=0$ increases exponentially.  However, if the subsystem length is sufficiently decreased by dephasing
$p>p^*>0$, the resistance transits from localized-exponential to ohmic-linear behavior. The critical dephasing probability $p^*$, which has been
calculated analytically in our recent work \cite{Zilly2012}, can be recognized approximately from \fig{6} as the $p$-value, where the resistance per
length diverges. Moreover, the critical dephasing probability is independent of the degree of momentum randomizing dephasing $p_r$
because the linear factor $(N-1)p(1-p_r)$ does not alter the fact that the resistance is exponentially increasing.

\fig{5} and the energy resolved transmission in \fig{7} show that the disorder-averaged resistance is always reduced by momentum conserving dephasing
$p_r=0$ because the fictitious dephasing reservoirs reduce the tunneling distance without causing additional resistance. Hence, the disorder-averaged
resistance is minimal if every bond is replaced by a momentum conserving dephasing reservoir. However, if the dephasing is momentum randomizing
$p_r>0$, the resistance depends crucially on the chain length, the disorder strength, the dephasing probabilities, and the energy. Compared to the
coherent resistance, the decoherent resistance may increase as well as decrease. Note that for a fixed disorder configuration, constructive
interference increases the transmission at some energies and hence momentum conserving dephasing reduces the transmission, see \fig{4}.

\section{Conclusions} \label{sec:conclusion}
This paper introduces a simple statistical model for the effects of dephasing on electron transport, which allows to adjust the degree of phase and
momentum randomization independently by using the dephasing probabilities $p$ and $p_r$ as well as the onsite disorder $\sg$. These parameters are
related to physical quantities, namely the phase coherence length \eq{lphi} and the mean free path \eq{lm}.

The resistance of a tight-binding chain \eq{Rav} indicates that momentum randomizing dephasing causes an additive resistance \eq{dR}, which increases
linearly with the chain length and both dephasing probabilities.

Studying a fixed disorder configuration, it has been shown in \fig{4} that only the oscillations in the transmission are averaged out if the dephasing
is momentum conserving, whereas the transmission itself is reduced if the dephasing is momentum randomizing.

The disorder-averaged resistance \eq{Ravd} is calculated very efficiently by considering the average number of coherent subsystems \eq{lav} for the
dephasing average and a recursion formula \eq{rdrec} for the disorder average of the coherent subsystems, which depends only on the variance of the
disorder distribution. Using this formula, it has been shown in \fig{6} that the dephasing-driven transition from localized-exponential to
ohmic-linear behavior is not affected by the degree of momentum randomizing dephasing because a linear factor does not alter the fact that the
resistance is increasing exponentially. However, the disorder-averaged resistance is always minimal if the dephasing is momentum conserving because
the tunneling distance is reduced without causing additional resistance.

Applying our model to homogeneous chains, it has been shown that the resistance is in between the ballistic and the ohmic regime, whereas for
disordered chains, the resistance is in between the exponential and ohmic regime. Hence, this model covers all three regimes, which have been defined
in the introduction.

\begin{acknowledgement}
  This work was supported by Deutsche Forschungsgemeinschaft under Grants No. GRK 1240 ``nanotronics'' and No. SPP 1386 ``nanostructured
  thermoelectric materials.'' O.U. acknowledges financial support of the J\'anos Bolyai Research Foundation of the Hungarian Academy of Sciences and
  the Hungarian NKTH-OTKA Grant No. CNK80991. We are very grateful to S.~Datta for useful discussions and helpful remarks.  T.S. enjoyed the
  hospitality of S.~Datta at the Purdue University and of O.~Ujs\'{a}ghy at the Budapest University of Technology and Economics.
\end{acknowledgement}


\end{document}